# Ferromagnetism and Ferroelectricity in epitaxial ultra-thin BiMnO$_3$ films

G. M. De Luca[1a)], D. Preziosi[2], F. Chiarella[1], R. Di Capua[1,3], S. Gariglio[4], S. Lettieri[1] and M. Salluzzo[1]

[1]CNR-SPIN, Complesso Monte S. Angelo, Via Cinthia, 80126 Napoli, Italy
[2]Max Planck Institute of Microstructure Physics, Weinberg 2, 06120 Halle, Germany
[3]Dipartimento di Fisica Università "Federico II" di Napoli, Complesso Monte S. Angelo, Via Cinthia, 80126 Napoli, Italy
[4]Département de Physique de la Matière Condensée, Université de Genève, 24 Quai Ernest Ansermet, 1211 Geneva 4, Switzerland

We studied the ferroelectric and ferromagnetic properties of compressive strained and unstrained BiMnO$_3$ thin films grown by rf-magnetron sputtering. BiMnO$_3$ samples exhibit a 2D cube-on-cube growth mode and a pseudo-cubic structure up to a thickness of 15 nm and of 25 nm when deposited on (001) SrTiO$_3$ and (110) DyScO$_3$, respectively. Above these thicknesses we observe a switching to a 3D island growth and a simultaneous structural change to a monoclinic structure characterized by a (00$l$) orientation of the monoclinic unit cell. While ferromagnetism is observed below T$_c$ $\cong$ 100 K for all samples, signatures of room temperature ferroelectricity were found only in the pseudo-cubic ultra-thin films, indicating a correlation between electronic and structural orders.

## I. Introduction

In the last years, the research on multiferroic materials has received a strong impulse because of their potential applications in the area of novel multifunctional devices as well as for the very interesting physics, involving the complex coexistence and coupling of ferroelectric and magnetic orders [1]. BiMnO$_3$ (BMO), which has been widely studied in form of polycrystalline samples, represents an interesting example, since it exhibits ferromagnetism below 100 K, with sizeable magnetic moments of the Mn$^{3+}$ ions comparable to the one of La$_{1-x}$Sr$_x$MnO$_3$ [2]. In addition, it is believed to be ferroelectric [2] with a monoclinic unit cell crystallizing in the non-centrosymmetric C2 space group ($a$ = 0.954nm, $b$ = 0.561nm, $c$ = 0.986nm, $\beta$ = 110.7°) [2]. The monoclinic unit cell is composed by the superposition of strongly distorted perovskite units with pseudocubic lattice parameter $a_{pc}$=3.95 Å. Early first-principles calculations [3] suggested proper ferroelectricity associated to the highly polarized 6$s^2$ character of Bi$^{3+}$ (lone pairs), as it happens in BiFeO$_3$. However, the occurence of ferroelectricity in this system has been challenged by a comparative analysis of electron and neutron diffraction data showing that polycrystalline BMO belongs instead to the centrosymmetric monoclinic C2/c space group, thus incompatible with ferroelectricity [4,5]. More recently, Solovyev *et al.* [6] proposed improper ferroelectricity of BiMnO$_3$ due to the competition between the ferromagnetic phase and a hidden antiferromagnetic phase, the latter associated to canted Mn spins associated to the 3D orbital order arrangement. The balance between these complex states could be tuned by applying external pressure [7] or epitaxial strain.

In this letter, we report on the growth of high quality BMO films on (001) SrTiO$_3$ (STO), with single TiO$_2$ termination, and (110) DyScO$_3$ (DSO) single-crystals. On STO, pseudo-cubic BMO films are characterized by a compressive bi-epitaxial strain (mismatch -1.1%), while the in-plane matching of BMO on (110) DSO is only slightly compressive (mismatch of -0.15%). The study of strained and unstrained films allowed an analysis of the role of the strain on the ferroelectric and ferromagnetic properties of this material.

**II. Samples fabrication and main characterizations**

The growth of BMO films is very difficult because of the high volatility of bismuth ions. The good control of the deposition parameters is extremely important to avoid the formation of competing phases, like BMO+$Mn_3O_4$ and BMO+$Bi_2O_{2.5}$, as observed by Lee et al. [8]. In particular, the temperature and the background oxygen during the physical vapor deposition play an important role on the realization of single phase epitaxial BMO films.

BMO samples were deposited by on-axis rf-magnetron sputtering. The crystal structure and the possible presence of spurious phases were verified by x-ray diffraction (Rigaku D-max and PANalytical X'Pert PRO). Finally the morphology of the samples deposited was analyzed by contact and non-contact Atomic Force microscopy (AFM, Park XE100). Differently from other reports [9-11], we used a stoichiometric target in order to minimize the formation of Bi-rich phases. Optimal growth conditions were a deposition pressure of 150 mTorr, composed by a mixture of Ar and $O_2$, with a ratio 10:1, a reduced rf power around 30 Watt, and substrate temperatures of 650 °C. We avoided any annealing in high oxygen pressure in order to prevent the stabilization of competing phases. Thus, after the growth the samples were cooled down to room temperature in the same pressure used during the deposition. The systematic study of the role of deposition parameters on the film properties confirmed that very low oxygen partial pressures during the deposition, very small deposition rates and not too high deposition temperatures are essential to avoid the formation of the secondary phases.

Another parameter playing a major role in the optimization of the BMO thin films was the target-substrate distance: in particular, we invariably observed the presence of $BiO_x$, $Mn_2O_3$ and (less frequently) $Mn_3O_4$ phases at target to substrate distances above 30 cm, independently on the thickness of the films.

BMO films exhibit a pseudo-cubic structure below a critical thickness of 25 nm for films deposited on (110) DSO and of 15 nm for films grown on (001) STO (Fig.1-2). Below these critical thicknesses, the samples show very flat and ordered surfaces (Fig. 1a and Fig. 1b) suggesting layer-by-layer or step-flow growth. Above the critical thickness, we observe a 2D-3D growth mode transition. Thick films are indeed characterized by multi-layered, and well ordered, 3D islands, often associated to screw-dislocations, as seen in Fig.1c and Fig. 1d. Simultaneously the orientation of the films changes from a pseudo-cubic structure characterized by the pseudo-cubic c-axis perpendicular to the substrate (Fig. 2b-2c), to a monoclinic structure, where the monoclinic $(00l)_m$ family of planes is parallel to the substrate surface. (Fig. 2a). This is schematically illustrated in Fig. 2d, where we show a plausible structural model of pseudo-cubic and monoclinic BMO deposited on STO. Below 100 nm, the samples have high crystalline quality, and are characterized by rocking curves, measured on the pseudo-cubic $(002)_{pc}$ and on the monoclinic $(003)_m$ reflections, having full widths at half maximum below 0.1°. Above 100 nm, cracks are created.

Concerning in plane epitaxy, x-ray diffraction maps of the pseudo-cubic (103) plane, show that thin BMO films deposited on STO are under compressive strain with the in-plane pseudocubic axes perfectly aligned to the STO lattice, and a *c*-axis of 3.99Å. The elongated *c*-axis is in perfect agreement with the expected value, assuming an ideally strained film and using the classical formula for biaxial strain:

$$\Delta c/c = [2\nu/(1-\nu)] * \Delta a/a. \quad (1)$$

Here, $\nu$ (=0.33) is the BMO Poisson coefficient [12], and $\Delta a$ and $\Delta c$ are the in-plane and out-of-plane lattice variations, respectively. Therefore, we can conclude that pseudo-cubic BMO thin films are characterized by a composition close to the stoichiometric one, since either oxygen or bismuth off-stoichiometry would substantially modify the structural parameters [13], contrary to our observations. On the other hand thick films show a monoclinic c-axis which is slightly shorter than the value expected for relaxed BMO.

**III. Results and Discussion**

Advanced magnetic characterizations have been performed by x-ray Magnetic Circular Dichroism (XMCD), at the ID08 beamline of the ESRF Synchrotron facility. Compared to standard magnetic characterizations, this technique allows a direct and element selective method to measure the magnetic moment of a specific ion in the compound, and it is especially suitable for thin films. First of all, an analysis of the x-ray absorption spectra on the optimized pseudocubic BMO films demonstrates that manganese is only in the $Mn^{3+}$ oxidation state. Thus we can exclude the presence of $Mn_3O_4$ ($Mn^{2+}$) spurious phases that could influence the magnetic behavior. The magnetic moment was obtained applying the sum rules [14], taking into account a correction factor calculated from a reference ferromagnetic $La_{1-x}Sr_xMnO_3$ sample. This correction factor is necessary, since it is known that the sum rules are not exact for Mn [15]. In fig. 3 we show the maximum value of the XMCD spectra normalized to the $L_3$ –edge, measured from 9 K to room temperature, in a magnetic field of 2 Tesla (zero field cooled). Both strained and unstrained BMO show a ferromagnetic transition around $T_c$=100K, which is similar to the well known bulk value.

The saturation magnetization, $M_{sat}$, determined from the XMCD data at 8 K and high field (>2 Tesla) is about 2.8 $\mu_B$/Mn and 3.2 $\mu_B$/Mn for films deposited on STO and DSO, respectively. These values are smaller than 4 $\mu_B$/Mn corresponding to fully aligned $Mn^{3+}$[4].

To investigate the ferroelectric properties of the films, we have grown BMO on metallic $LaNiO_3$ buffer layers (deposited on STO or on Nb-doped STO by off-axis rf-sputtering at University of Geneva [16].

While the in plane resistivity is very large even at room temperature, in particular for thin pseudo-cubic films, standard ferroelectric test measurements in a sandwich configuration failed due to the high leakage.

For these reasons, evidences of ferroelectric behavior were sought by piezo-force microscopy (PFM) measurements at room temperature. In fig. 4a we show an examples of PFM measurement on pseudo-cubic BMO film deposited on STO substrate. Complex geometrical patterns can be effectively written by using the AFM probe and applying different negative and positive dc-biases (± 8 V bright and dark regions in the amplitude of PFM response, see fig. 4a). We observe all the typical signatures of ferroelectric domain switching. In particular, 180° switching of the phase among regions with reversed polarization and phase/ amplitude vs. tip voltage loops typical of ferroelectric materials (Fig. 4b) [17]. Thus, the PFM response in both amplitude and phase, can be related to the different signs of a component of the polarization perpendicular to the film surface, However, these results represent a necessary, but not a sufficient proof in favor of room temperature ferroelectricity in our films. As discussed in [18], there are two possible alternative explanations: 1) simple charging of the surface; 2) ionic displacements of the oxygen ions towards the surface in a direction defined by the applied electric field. The latter effect is particularly subtle, since the movement of the oxygen ions from their original positions gives rise to a false piezostrictive effect [18].

For this reason, we investigated the time dependence of the amplitude of the piezo-response signal. It is well known, that the polarization of ferroelectric materials is characterized a loss which is, on time scales of days, typically of few percents [19]. This quantity can be evaluated by measuring the PFM amplitude signal as function of the time. As shown in Fig. 4c, the PFM amplitude measured on a pseudo-cubic 10 nm-thick BMO film deposited on DSO decays, at first approximation, by following a simple logarithmic law, also typical of ferroelectric materials [20]:

$$\Delta P(t) = \Delta P_0 - m \log[t/t_0]. \quad (2)$$

In eq. (2), $t_0$ is the characteristic time above which the polarization starts to decay more rapidly and linearly, and $\Delta P_0$ is the retention polarization at $t=t_0$, and m is the decay rate. In our compound, we find a retention polarization equal to 50% of the initial value (for both plus and minus polarizations). However, the initial decay may indicate an important contribution from trapped charges at the surface and subsequent relaxation with the sample remaining in air after long time. However, after this decay the signal remains stable for days. We mention that on thick films, characterized by a 3D growth mode and a different orientation, we were never able to detect a stable PFM, ferroelectric-like, signal. Thus, these data, even if are not conclusive, suggest that pseudo-cubic thin films are indeed ferroelectric also at room temperature.

It is worth noting that the main differences between thin and thick BMO films concern their structure and their strain state. Thin films are epitaxially strained and possed a pseudo-cubic lattice. Unfortunately we cannot firmly establish, on the base of our structural data, if thin films have a structure belongin to centrosymmetric or not centrosymmetric space groups. However, the structural distortions induced by the substrate are likely sufficient to make these samples substan-tially different from relaxed monoclinic BMO. On the other hand, thick films are relaxed and consequently have structural properties closer to bulk BMO. Thus, we attribute the ferroelectric behavior observed in strained pseudo-cubic BMO films to their structure.

More investigations, using advanced spectroscopy, are then required to establish whether crystalline pseudocubic BMO films are also multiferroic.

In conclusion, we have found the right deposition condition to grow high quality single phase and extremely flat, ferromagnetic BMO thin films, with single crystal orientation using rf-magnetron sputtering. Piezo-force microscopy reveals that pseudo-cubic thin films are characterized by a ferroelectric-like switching and that the structural order plays an important role in the ferroelectric behavior of these films. These results are promising in view of future investigation of multiferroic effects in strongly ferromagnetic $BiMnO_3$.

**Acknowledgments**

The authors are grateful to P. Zubko for the characterization of $BiMnO_3/SrTiO_3$ films and to V. Sessi and N. B. Brookes for the support during the x-ray spectroscopy measurements. The research leading to these results has received funding from the European Union Seventh Framework Program (FP7/2007-2013) under Grant Agreement No. 264098-MAMA.

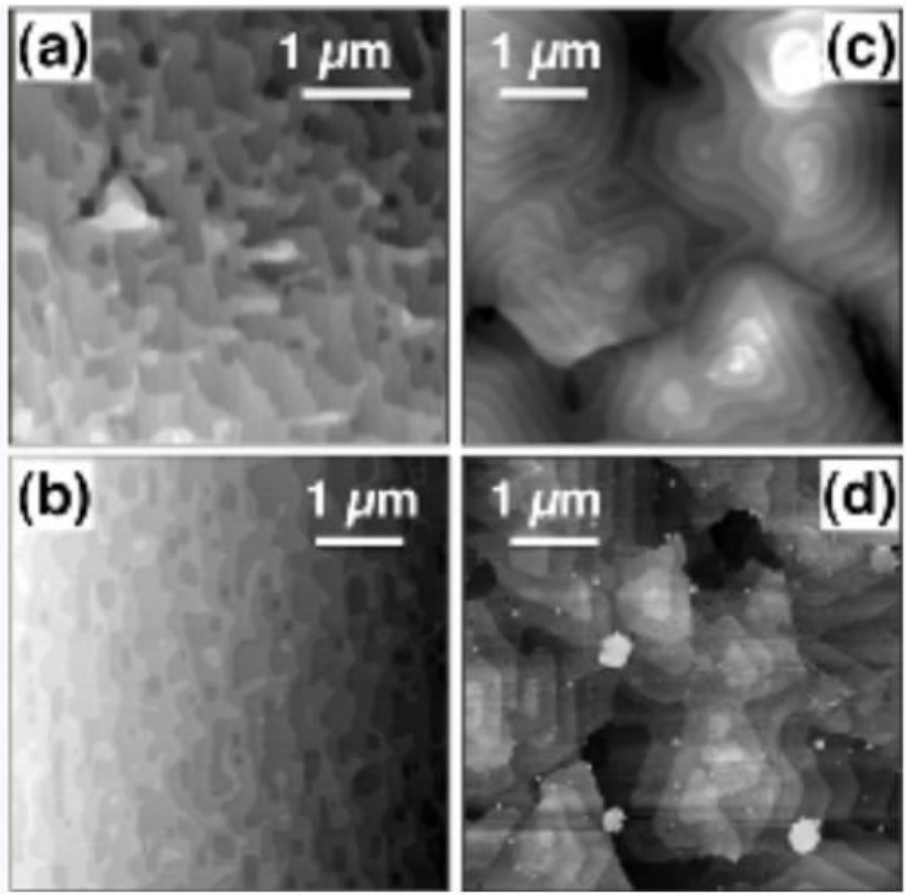

Figure 1: AFM image of different thickness BMO films grown on STO and DSO substrates: a) 10 nm-thick BMO on STO; b) 10 nm-thick BMO on DSO; c) 60 nm-thick BMO on STO; d) 60 nm-thick BMO on DSO.

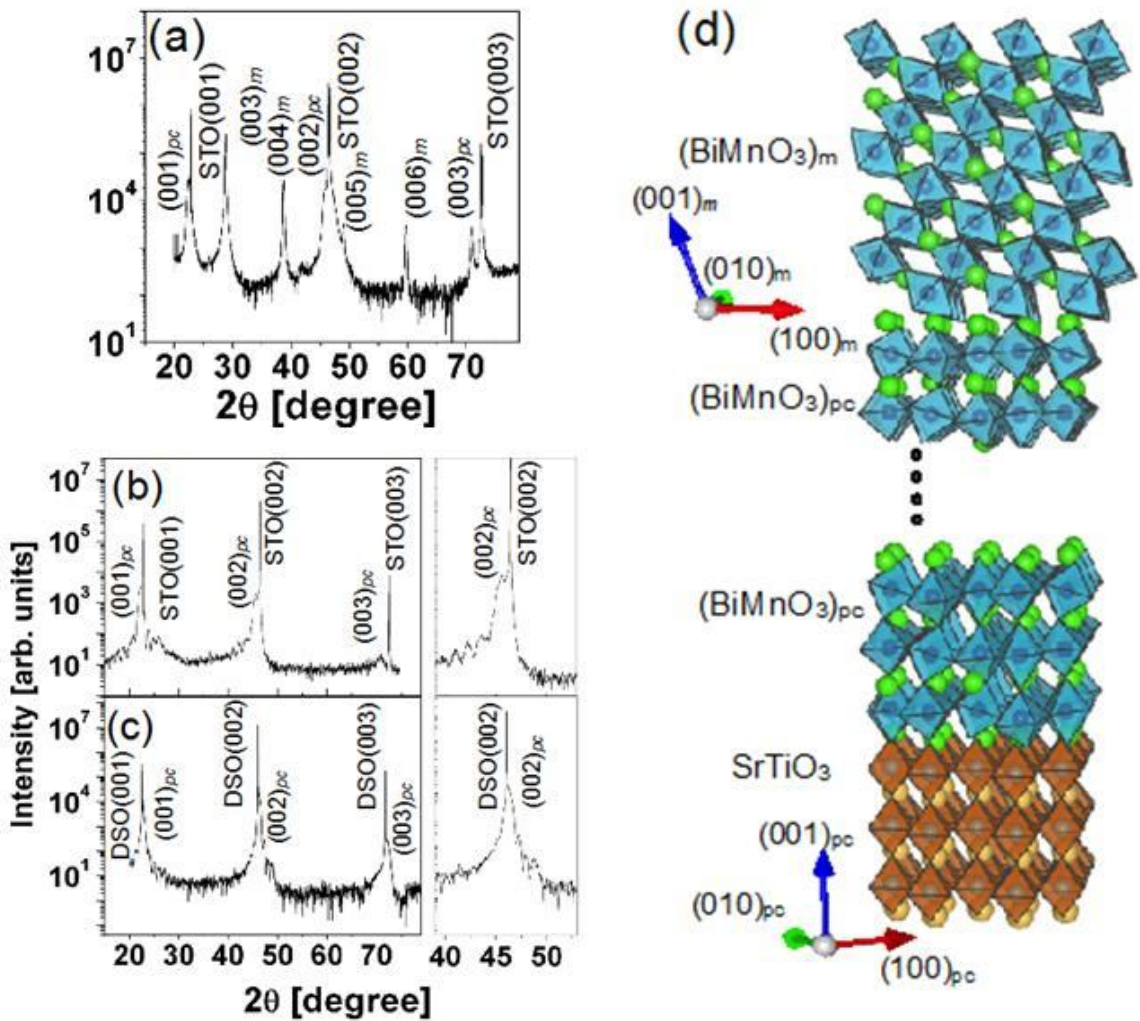

Figure 2: θ-2θ x-ray diffraction on (a) 60 nm.thick BiMnO3 film deposited on (001) STO, (b) 10nm-thick BMO grown on STO and c) 10nm-thick BMO grown on DSO substrates. Finite size effect oscillations can be observed on the pseudo-cubic films in b) and c). d) a sketch of a plausible structural model of pseudo-cubic and monoclinic BMO deposited on STO.

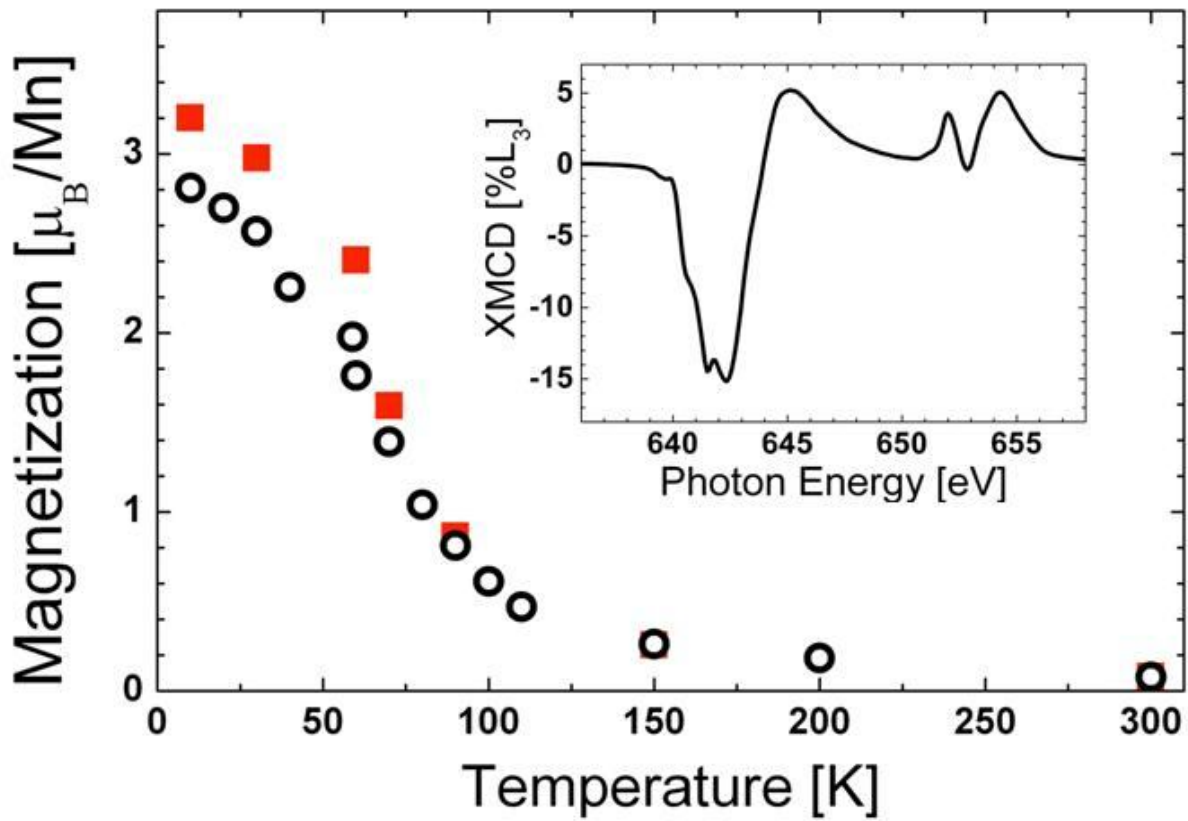

Figure 3: Temperature dependence of the magnetization obtained using sum rules applied to XMCD spectra acquired on 10 nm-thick BMO films deposited on STO (black open circle) and DSO (full red square). The data have been acquired warming up the samples from 8 K to 300 K in a field of 2 Tesla, almost parallel to the surface. In The inset shows an example of Mn $L_{2,3}$ edge XMCD spectra (10K, 3 Tesla).

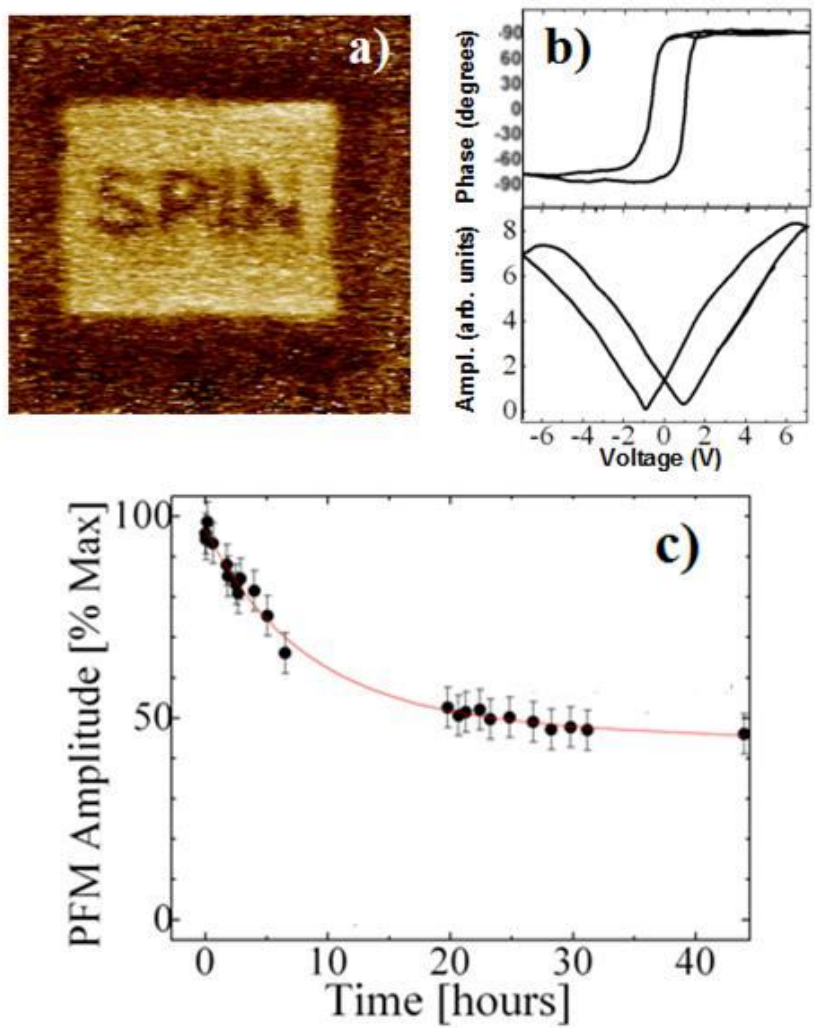

Figure 4: PFM amplitude image obtained on 10nm-thick BMO films on STO, showing a complex pattern created by applying positive (bright regions) and negative (dark regions) tip bias voltages of ±8 V. b) Phase vs bias voltage and c) amplitude vs bias voltage hysteresis loops obtained on a given location of the sample. d) PFM amplitude as function of the time acquired on a 10nm-thick BMO thin film grown on DSO. the red line is a fit using eq. (2).